# Dormant entanglement that can be activated or destroyed by the basis choice of measurements on an external system

Zixuan Hu and Sabre Kais*

*Department of Chemistry, Department of Physics, and Purdue Quantum Science and Engineering Institute, Purdue University, West Lafayette, IN 47907, United States*
*Email:* kais@purdue.edu

Abstract: We propose a new form of entanglement called the dormant entanglement that can be activated or destroyed by the basis choice of measurements on an external system. The dormant entanglement without activation cannot be used as a quantum communication resource and has reduced correlation as compared to the Bell states. A particular form of the dormant entanglement is so weak that, without activation, no correlation in any basis can be observed between the entangled qubits. The dormant entanglement showcases a unique quantum behavior that the physical description of a local system remains incomplete until the information on all external systems entangled with the local system becomes available. For a potential application, we propose an $n$-party collective quantum communication channel that allows any 2 out of the $n$ parties to activate an entanglement pair with the complete consensus of all other parties.

## 1. Introduction

Quantum information science and quantum communication have received intensive research interests in the last 30 years, with sophisticated theories and state-of-the-art technologies emerging one after another [1-31]. Entanglement is an important phenomenon and valuable resource in quantum information science and quantum communication. Entanglement found in model systems such as the Bell states and the GHZ (Greenberger–Horne–Zeilinger) state demonstrates quantum-only correlation behaviors when the qubits are measured in different bases [32, 33]. Exploiting the non-locality of entanglement, quantum algorithms have been designed to realize novel quantum communication schemes such as quantum teleportation [34, 35] and quantum key distribution [36-41].

In this work, we propose a new form of entanglement called the dormant entanglement. When we prepare the quantum state in a particular structure, such that two qubits are entangled with each other and there is an external system entangled with the 2-qubit subsystem, the entanglement of the subsystem can stay dormant until getting activated or destroyed, depending on the basis in which the external system is measured. When the entanglement is dormant, the 2-qubit subsystem behaves like an unentangled state, has reduced correlation between the qubits, and cannot be used as a quantum communication channel. When the dormant entanglement is activated, the 2-qubit subsystem behaves like an entangled Bell state and can be used as a quantum communication channel. We then proceed to study the correlation behavior of the dormant entanglement state quantitatively with the CHSH-test (Clauser-Horne-Shimony-Holt) [42]. We find that by adding an additional "lock qubit" to the 3-qubit dormant entanglement state, a "weakest form" of entanglement can be created – weakest in the sense of having no correlation in any basis and zero value for the CHSH-test.

The fact that dormant entanglement can be activated or destroyed by the basis choice of measurements on an external system has deep implication for understanding the qubit quantum space: it showcases a unique quantum behavior that the physical description of a local system remains incomplete until information on all external systems entangled with the local system becomes available. In addition, the dormant entanglement may have potential applications in quantum communication: we propose an *n*-party collective quantum communication channel that allows any 2 out of the *n* parties to activate an entanglement pair by the complete consensus of all other parties. Compared to establishing point-to-point entanglement channel between any 2 qubits, the collective quantum channel reduces the required number of qubits.

## 2. The dormant entanglement

Consider the 3-qubit quantum state:

$$\left|\psi^{(3)}\right\rangle = \frac{1}{2}\Big[\big(\left|00\right\rangle_{12} + \left|11\right\rangle_{12}\big)\left|0\right\rangle_3 + \big(\left|01\right\rangle_{12} + \left|10\right\rangle_{12}\big)\left|1\right\rangle_3\Big] \tag{1}$$

where the superscript of $\left|\psi^{(3)}\right\rangle$ indicates the state has 3 qubits, and the subscript of e.g. $\left|01\right\rangle_{12}$ indicates in this basis state the 1st qubit $q_1 = \left|0\right\rangle$ and the 2nd qubit $q_2 = \left|1\right\rangle$. $\left|\psi^{(3)}\right\rangle$ can be created with the simple circuit shown in Figure 1:

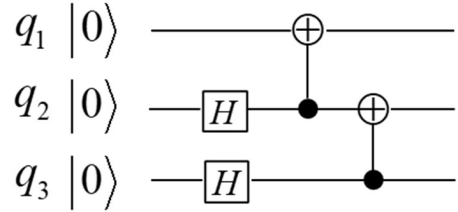

Figure 1. The quantum circuit for creating the 3-qubit dormant entanglement state $\left|\psi^{(3)}\right\rangle$.

Firstly we show $q_1$ and $q_2$ are entangled. If we measure $q_3$ in the current basis, then if $q_3 = \left|0\right\rangle$ we have $\left|\varphi_1\right\rangle = \frac{1}{\sqrt{2}}\big(\left|00\right\rangle_{12} + \left|11\right\rangle_{12}\big)$, and if $q_3 = \left|1\right\rangle$ we have $\left|\varphi_2\right\rangle = \frac{1}{\sqrt{2}}\big(\left|01\right\rangle_{12} + \left|10\right\rangle_{12}\big)$. Clearly both $\left|\varphi_1\right\rangle = \frac{1}{\sqrt{2}}\big(\left|00\right\rangle_{12} + \left|11\right\rangle_{12}\big)$ and $\left|\varphi_2\right\rangle = \frac{1}{\sqrt{2}}\big(\left|01\right\rangle_{12} + \left|10\right\rangle_{12}\big)$ are entangled states, and thus if $q_1$ and $q_2$ were not entangled in the original $\left|\psi^{(3)}\right\rangle$, then we would have created the $q_1$ - $q_2$

entanglement by a local measurement on $q_3$ not involving $q_1$ or $q_2$ – which is impossible and therefore $q_1$ and $q_2$ are already entangled in $\left|\psi^{(3)}\right\rangle$.

Next however, we show $q_1$ and $q_2$ can also be unentangled when $q_3$ is measured in a different basis. If we rotate $q_3$ by a Hadamard gate, then $\left|\psi^{(3)}\right\rangle$ becomes:

$$
\begin{aligned}
H_3\left|\psi^{(3)}\right\rangle &= \frac{1}{2\sqrt{2}}\Big[\big(\left|00\right\rangle_{12}+\left|11\right\rangle_{12}\big)\big(\left|0\right\rangle_3+\left|1\right\rangle_3\big)+\big(\left|01\right\rangle_{12}+\left|10\right\rangle_{12}\big)\big(\left|0\right\rangle_3-\left|1\right\rangle_3\big)\Big] \\
&= \frac{1}{2\sqrt{2}}\Big[\big(\left|0\right\rangle_1+\left|1\right\rangle_1\big)\big(\left|0\right\rangle_2+\left|1\right\rangle_2\big)\left|0\right\rangle_3+\big(\left|0\right\rangle_1-\left|1\right\rangle_1\big)\big(\left|0\right\rangle_2-\left|1\right\rangle_2\big)\left|1\right\rangle_3\Big] \\
&= \frac{1}{\sqrt{2}}\Big[\left|++\right\rangle_{12}\left|0\right\rangle_3+\left|--\right\rangle_{12}\left|1\right\rangle_3\Big]
\end{aligned}
\tag{2}
$$

$H_3 = I \otimes I \otimes H$ is the Hadamard gate applied on $q_3$

$$
\left|+\right\rangle = \frac{1}{\sqrt{2}}\big(\left|0\right\rangle+\left|1\right\rangle\big), \quad \left|-\right\rangle = \frac{1}{\sqrt{2}}\big(\left|0\right\rangle-\left|1\right\rangle\big)
$$

In Equation (2), if we measure $q_3 = \left|0\right\rangle$, we have $\left|\phi_1\right\rangle = \left|++\right\rangle_{12}$, and if $q_3 = \left|1\right\rangle$ we have $\left|\phi_2\right\rangle = \left|--\right\rangle_{12}$: clearly both $\left|\phi_1\right\rangle$ and $\left|\phi_2\right\rangle$ are unentangled product states.

So there is an apparent paradox that $q_1$ and $q_2$ are entangled or unentangled, depending on the basis choice of the external qubit $q_3$ – rotating $q_3$ by a Hadamard and measuring it is effectively just measuring $q_3$ in the $\{\left|+\right\rangle, \left|-\right\rangle\}$ basis. The key to resolve the paradox is to recognize the fact that $q_1$ and $q_2$ are unentangled in their own subsystem and are entangled when the total system including $q_3$ is considered. Deriving from the 3$^{\text{rd}}$ line of Equation (2) where $q_3$ has been rotated by a Hadamard gate, the $q_1$-$q_2$ subsystem can be described by the reduced density matrix:

$$
\rho_{12} = \frac{1}{2}\big(\left|++\right\rangle\left\langle++\right| + \left|--\right\rangle\left\langle--\right|\big)
\tag{3}
$$

In Equation (3) $\rho_{12}$ can be written as a mixture of two unentangled product states, and thus $\rho_{12}$ itself is also unentangled [43, 44] – this means if we focus on the $q_1$-$q_2$ subsystem described by $\rho_{12}$, it is indeed unentangled. However, now deriving from Equation (1) where $q_3$ is in the original basis, the reduced density matrix of the $q_1$-$q_2$ subsystem is:

$$\rho'_{12} = \frac{1}{2}\left(|\varphi_1\rangle\langle\varphi_1| + |\varphi_2\rangle\langle\varphi_2|\right)$$

$$|\varphi_1\rangle = \frac{1}{\sqrt{2}}\left(|00\rangle_{12} + |11\rangle_{12}\right), \qquad |\varphi_2\rangle = \frac{1}{\sqrt{2}}\left(|01\rangle_{12} + |10\rangle_{12}\right) \tag{4}$$

$\rho'_{12}$ in Equation (4) looks distinct from $\rho_{12}$ in Equation (3) because $\rho'_{12}$ is written as a mixture of two entangled states. The physical description of the $q_1$-$q_2$ subsystem should not depend on the basis choice of the external $q_3$, and we indeed find that $\rho'_{12} = \rho_{12}$ mathematically: this means regardless of the basis choice of the external $q_3$, any local measurements involving only the $q_1$-$q_2$ subsystem should yield indistinguishable results for $\rho'_{12}$ and $\rho_{12}$.

However, the equivalence of $\rho'_{12}$ and $\rho_{12}$ stops when we include $q_3$ into the consideration. If $q_3$ is measured in the original basis in Equation (1), the $q_1$-$q_2$ subsystem is in the mixture of $\rho'_{12}$, but this mixture can be separated into the entangled states $|\varphi_1\rangle = \frac{1}{\sqrt{2}}\left(|00\rangle_{12} + |11\rangle_{12}\right)$ for $q_3 = |0\rangle$ and $|\varphi_2\rangle = \frac{1}{\sqrt{2}}\left(|01\rangle_{12} + |10\rangle_{12}\right)$ for $q_3 = |1\rangle$: **this means the entanglement of $q_1$ and $q_2$ can be activated by the measurement result of $q_3$ in the original basis**. On the other hand, if $q_3$ is measured in the rotated basis in Equation (2), the $q_1$-$q_2$ subsystem is in the mixture of $\rho_{12}$ in Equation (3), and this mixture can only be separated into the unentangled states $|++\rangle$ for $q_3 = |0\rangle$ and $|--\rangle$ for $q_3 = |1\rangle$: **this means the entanglement of $q_1$ and $q_2$ can be destroyed by measuring $q_3$ in the rotated basis**. In the following we define the concept of dormant entanglement:

**Definition:** In the 3-qubit state $|\psi^{(3)}\rangle = \frac{1}{2}\left[\left(|00\rangle_{12} + |11\rangle_{12}\right)|0\rangle_3 + \left(|01\rangle_{12} + |10\rangle_{12}\right)|1\rangle_3\right]$, the entanglement of the $q_1$-$q_2$ subsystem can be activated by measuring $q_3$ in the original $\{|0\rangle, |1\rangle\}$ basis or be destroyed by measuring $q_3$ in the Hadamard-rotated $\{|+\rangle, |-\rangle\}$ basis. In such a situation we define the $q_1$-$q_2$ entanglement as the **dormant entanglement**, and $q_3$ as its **controller**.

Below we discuss several interesting properties of the dormant entanglement:

1. To activate the dormant entanglement of $q_1$ and $q_2$, the controller $q_3$ must be measured in the original $\{|0\rangle, |1\rangle\}$ basis, and the result must be communicated to the $q_1$-$q_2$ subsystem to separate $\rho'_{12}$ into the entangled states. If $q_3$ is measured but the result is lost at any stage before reaching the $q_1$-$q_2$ subsystem, $q_1$-$q_2$ will stay in $\rho'_{12} = \rho_{12}$, which is itself

unentangled without separation. This fact ensures that the speed of information transfer from $q_3$ to the $q_1$-$q_2$ subsystem is bounded by the speed of light (i.e. no violation of special relativity).

2. Different from the activation above, to destroy the dormant entanglement of $q_1$ and $q_2$, one only needs to measure the controller $q_3$ in the Hadamard-rotated $\{|+\rangle, |-\rangle\}$ basis, and the result of the measurement does not need to be recorded or communicated. In addition, the destruction of the dormant entanglement is permanent in the sense that any possibility of activating it is irreversibly lost.

3. By the quantum no-communication theorem [34], without any information from $q_3$, the $q_1$-$q_2$ subsystem cannot by any local means tell if $q_3$ has been measured in whatever basis. In fact, if $q_1$ and $q_2$ are not aware of the existence of $q_3$ in the first place, they cannot by any local means tell if $q_3$ exists. If $q_3$ does not exist, the $q_1$-$q_2$ system can also be in the density matrix of $\rho'_{12} = \rho_{12}$ but with no means of activating the entanglement. Consequently it is fundamentally impossible for $q_1$ and $q_2$ to tell if their system is "truly unentangled" without an external $q_3$, or "dormantly entangled" such that it can be activated by an external $q_3$. **This showcases a unique quantum behavior that the physical description of a local system remains incomplete until the information on all external systems entangled with the local system becomes available.** In other words, the information on the external system $q_3$ – whether it exists or when it does how it is measured – is indeed physical, because it can be used to change the physical description of the $q_1$-$q_2$ system, from being untangled to entangled.

4. Without activation, the dormant entanglement cannot be used as a quantum communication channel. Indeed, without any information or interaction from $q_3$, the $q_1$-$q_2$ subsystem is in the state of $\rho'_{12} = \rho_{12}$ which is unentangled, so it cannot be used as a resource for quantum communication.

### 3. The qubit equivalence and the collective quantum communication channel

Having defined the dormant entanglement and discussed its properties, now we go back to $|\psi^{(3)}\rangle$ in Equation (1) and add more qubits to it by repeating the procedure in Figure 1, as shown in Figure 2:

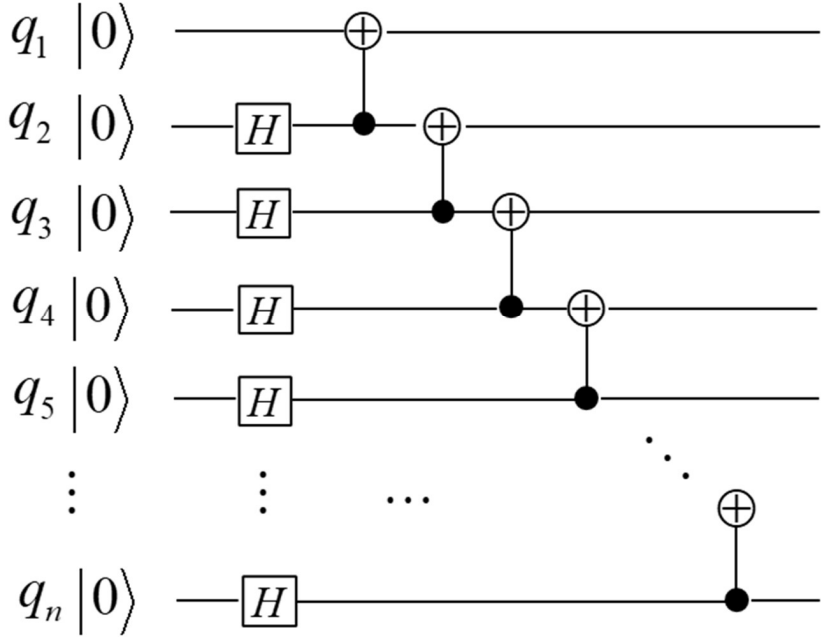

Figure 2. The quantum circuit for creating the *n*-qubit dormant entanglement state $\left|\psi^{(n)}\right\rangle$.

By the procedure in Figure 2 we can indefinitely add qubits one-by-one to the growing entanglement structure and create the *n*-qubit state $\left|\psi^{(n)}\right\rangle$ which has the same structure as $\left|\psi^{(3)}\right\rangle$. For example, the 5-qubit state $\left|\psi^{(5)}\right\rangle$ looks like:

$$\left|\psi^{(5)}\right\rangle = \frac{1}{4}\left\{ \begin{bmatrix} \left(\left|00\right\rangle_{12}+\left|11\right\rangle_{12}\right)\left|0\right\rangle_3 + \left(\left|01\right\rangle_{12}+\left|10\right\rangle_{12}\right)\left|1\right\rangle_3 \end{bmatrix}\left|0\right\rangle_4 \\ + \begin{bmatrix} \left(\left|00\right\rangle_{12}+\left|11\right\rangle_{12}\right)\left|1\right\rangle_3 + \left(\left|01\right\rangle_{12}+\left|10\right\rangle_{12}\right)\left|0\right\rangle_3 \end{bmatrix}\left|1\right\rangle_4 \right\}\left|0\right\rangle_5$$
$$+ \frac{1}{4}\left\{ \begin{bmatrix} \left(\left|00\right\rangle_{12}+\left|11\right\rangle_{12}\right)\left|0\right\rangle_3 + \left(\left|01\right\rangle_{12}+\left|10\right\rangle_{12}\right)\left|1\right\rangle_3 \end{bmatrix}\left|1\right\rangle_4 \\ + \begin{bmatrix} \left(\left|00\right\rangle_{12}+\left|11\right\rangle_{12}\right)\left|1\right\rangle_3 + \left(\left|01\right\rangle_{12}+\left|10\right\rangle_{12}\right)\left|0\right\rangle_3 \end{bmatrix}\left|0\right\rangle_4 \right\}\left|1\right\rangle_5 \tag{5}$$

For Equation (5) it is easy to verify that the $q_1$-$q_2$ subsystem of $\left|\psi^{(5)}\right\rangle$ (or in general any other $\left|\psi^{(n)}\right\rangle$) are still in the state of $\rho'_{12}=\rho_{12}$, but now to activate the $q_1$-$q_2$ entanglement we will need to measure all the other qubits – i.e. $q_3$, $q_4$, and $q_5$ – in the current $\left\{\left|0\right\rangle,\left|1\right\rangle\right\}$ basis and communicate the results to the $q_1$-$q_2$ subsystem. On the other hand if any one of the other qubits is measured in the $\left\{\left|+\right\rangle,\left|-\right\rangle\right\}$ basis then the $q_1$-$q_2$ entanglement is destroyed. Here we can revise the definition of the dormant entanglement to have multiple qubits as the controllers, and all the controllers must reach complete consensus for measuring in the correct basis in order to activate the dormant entanglement of the $q_1$-$q_2$ subsystem.

Now an important property of $\left|\psi^{(n)}\right\rangle$ is that its structure remains unchanged upon any permutation of the qubits. To prove this we first consider $\left|\psi^{(3)}\right\rangle$ as in Equation (1), and after some basic algebra find its structure is unchanged upon any permutation of the qubits:

$$
\begin{aligned}
\left|\psi^{(3)}\right\rangle &= \frac{1}{2}\Big[\big(\left|00\right\rangle_{12}+\left|11\right\rangle_{12}\big)\left|0\right\rangle_3 + \big(\left|01\right\rangle_{12}+\left|10\right\rangle_{12}\big)\left|1\right\rangle_3\Big] \\
&= \frac{1}{2}\Big[\big(\left|00\right\rangle_{13}+\left|11\right\rangle_{13}\big)\left|0\right\rangle_2 + \big(\left|01\right\rangle_{13}+\left|10\right\rangle_{13}\big)\left|1\right\rangle_2\Big] \\
&= \frac{1}{2}\Big[\big(\left|00\right\rangle_{23}+\left|11\right\rangle_{23}\big)\left|0\right\rangle_1 + \big(\left|01\right\rangle_{23}+\left|10\right\rangle_{23}\big)\left|1\right\rangle_1\Big] \\
&= \ \dots \text{ any qubit permutation}
\end{aligned}
\tag{6}
$$

Now by mathematical induction (details of the proof can be found in the Supplementary Information (SI)), if the structure is unchanged upon qubit permutation for $\left|\psi^{(k)}\right\rangle$, then this is also true for $\left|\psi^{(k+1)}\right\rangle$ because $\left|\psi^{(k+1)}\right\rangle$ is created by adding one more qubit to $\left|\psi^{(k)}\right\rangle$ while preserving the structure. Consequently we have the structure of any $\left|\psi^{(n)}\right\rangle$ remains unchanged upon any permutation of the qubits. In other words, all qubits in $\left|\psi^{(n)}\right\rangle$ are equivalent, therefore any two qubits are in dormant entanglement with the other qubits being its controllers.

The qubit equivalence of $\left|\psi^{(n)}\right\rangle$ allows us to propose a collective quantum communication channel as shown in Figure 3.

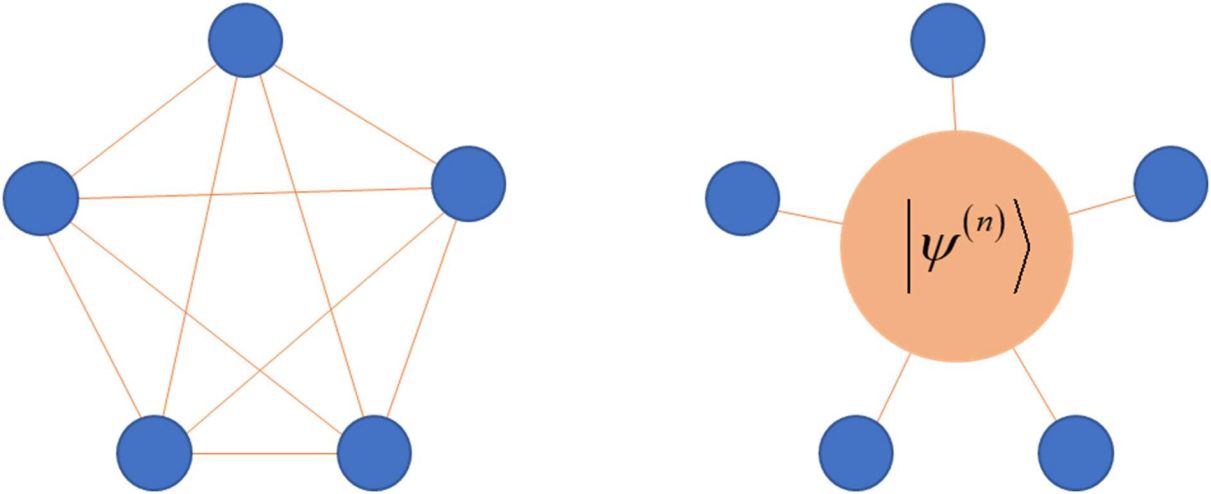

Figure 3. The collective quantum communication channel that allows any two parties to activate an entanglement quantum channel with the complete consensus of all other parties. On the left is the point-to-point structure where every pair of parties shares an entanglement pair, and totally $2 \cdot C(5,2) = 5(5-1) = 20$ qubits are required. On the right is the collective quantum channel based on the $n$-qubit dormant entanglement state $\left| \psi^{(n)} \right\rangle$, and totally 5 qubits are required for one pair of parties to communicate.

Suppose we want to establish a quantum communication network for $n$ parties. Suppose we do not know in advance which two parties will need to communicate, and we want to avoid creating entanglements at the time of communication – i.e. we can only perform non-local quantum operations in advance and only perform local quantum operations at the time of communication. Then we need to ensure every two parties are connected by an entanglement channel that can allow quantum teleportation [34, 35], quantum key distribution [36-41], or other quantum communication tasks. If we create a point-to-point entanglement for every two parties, we need to use $C(n,2) = \frac{1}{2} n(n-1)$ entanglement pairs, which are $n(n-1)$ qubits. However, if we create the $n$-qubit dormant entanglement structure of $\left| \psi^{(n)} \right\rangle$, and assign one qubit to each party, then by the equivalence of qubits, when at the time of communication any two parties need to use an entanglement quantum channel, we just need to activate the dormant entanglement of their qubits by local measurements on the controllers and communication of the results. Compared to the point-to-point structure, we reduce the required number of qubits from $n(n-1)$ to $n$. Of course, if $k$ pairs of parties need to communicate, we will need to create $k$ copies of $\left| \psi^{(n)} \right\rangle$, and the required number of qubits increases to $kn$, so if $k \ll n$ the collective quantum channel based on $\left| \psi^{(n)} \right\rangle$ will have an advantage over the point-to-point structure. An additional feature of the $\left| \psi^{(n)} \right\rangle$ based structure is the dormant entanglement channel between any two parties can only be activated by the complete consensus of all other parties, which is distinct from the point-to-point structure where any two parties can communicate without any input from other parties.

## 4. The weakest entanglement with no correlation in any basis

A distinguishing feature of entanglement is the correlation of two qubits in different bases. Classical systems can also have correlations, but the Bell states can have perfect correlations in both the original $\left\{ \left| 0 \right\rangle, \left| 1 \right\rangle \right\}$ basis and the Hadamard-rotated $\left\{ \left| + \right\rangle, \left| - \right\rangle \right\}$ basis, for example:

$$\left| \varphi_1 \right\rangle = \frac{1}{\sqrt{2}} \left( \left| 00 \right\rangle_{12} + \left| 11 \right\rangle_{12} \right) = \frac{1}{\sqrt{2}} \left( \left| ++ \right\rangle_{12} + \left| -- \right\rangle_{12} \right)$$

$$\left| \varphi_2 \right\rangle = \frac{1}{\sqrt{2}} \left( \left| 01 \right\rangle_{12} + \left| 10 \right\rangle_{12} \right) = \frac{1}{\sqrt{2}} \left( \left| ++ \right\rangle_{12} - \left| -- \right\rangle_{12} \right)$$

$$(7)$$

A quantitative way to represent the unique quantum correlation in the Bell states is the CHSH-test [42]. In the CHSH-test, each qubit of a Bell state is measured in two different bases and the correlation of the Bell state is represented by a sum of four observables as in Equation (8):

$q_1$ measurements:   $A_0 = \sigma_z$,   $A_1 = \sigma_x$,

$q_2$ measurements:   $B_0 = -\dfrac{1}{\sqrt{2}}(\sigma_x + \sigma_z)$,   $B_1 = \dfrac{1}{\sqrt{2}}(\sigma_x - \sigma_z)$   (8)

Sum of observables:   $S = \pm\langle A_0 \otimes B_0 \rangle \pm \langle A_0 \otimes B_1 \rangle \pm \langle A_1 \otimes B_0 \rangle \pm \langle A_1 \otimes B_1 \rangle$

The sum $S$ in Equation (8) can be maximized by choosing a particular combination of the signs (not all sign combinations are allowed, see the SI for details) before the observables, given the particular Bell state being evaluated, for example:

For   $|\varphi_1\rangle = \dfrac{1}{\sqrt{2}}\left(|00\rangle_{12} + |11\rangle_{12}\right)$,

$S_{\max}\left(|\varphi_1\rangle\right) = -\langle A_0 \otimes B_0 \rangle - \langle A_0 \otimes B_1 \rangle - \langle A_1 \otimes B_0 \rangle + \langle A_1 \otimes B_1 \rangle = 2\sqrt{2}$

(9)

For   $|\varphi_2\rangle = \dfrac{1}{\sqrt{2}}\left(|01\rangle_{12} + |10\rangle_{12}\right)$,

$S_{\max}\left(|\varphi_2\rangle\right) = +\langle A_0 \otimes B_0 \rangle + \langle A_0 \otimes B_1 \rangle - \langle A_1 \otimes B_0 \rangle + \langle A_1 \otimes B_1 \rangle = 2\sqrt{2}$

Note that for all the four Bell states the maximum sum $S_{\max}$ is always $2\sqrt{2}$, which is greater than the classical limit of 2, and thus the Bell states possess unique quantum correlations.

Now what happens if we study the correlation in different bases for the dormant entanglement of $q_1$ and $q_2$ in $|\psi^{(n)}\rangle$? Take $|\psi^{(3)}\rangle$ in Equation (1) as an example, clearly $q_1$ and $q_2$ have no correlation when measured in the $\{|0\rangle, |1\rangle\}$ basis, because without measuring $q_1$, the probability of measuring $q_2 = |0\rangle$ is $p(q_2 = |0\rangle) = \dfrac{1}{2}$; when $q_1$ is measured to be either $|0\rangle$ or $|1\rangle$, then the conditional probability of measuring $q_2 = |0\rangle$ given $q_1$'s result is:

$$p(q_2 = |0\rangle \,|\, q_1 = |0\rangle) \;=\; p(q_2 = |0\rangle \,|\, q_1 = |1\rangle) \;=\; \dfrac{1}{2} \;=\; p(q_2 = |0\rangle) \qquad (10)$$

We see in Equation (10) that the probability of measuring $q_2 = |0\rangle$ is always $\dfrac{1}{2}$, regardless if $q_1$ is measured or not. This means $q_1$ and $q_2$ are not correlated when measured in the $\{|0\rangle, |1\rangle\}$ basis. On the other hand, if we apply the Hadamard gate to both $q_1$ and $q_2$, and then measure the qubits – i.e. measure in the $\{|+\rangle, |-\rangle\}$ basis, we have:

$$H_1 H_2 \left| \psi^{(3)} \right\rangle = \frac{1}{4} \left( \begin{array}{c} \left[ \left( \left|0\right\rangle_1 + \left|1\right\rangle_1 \right) \left( \left|0\right\rangle_2 + \left|1\right\rangle_2 \right) + \left( \left|0\right\rangle_1 - \left|1\right\rangle_1 \right) \left( \left|0\right\rangle_2 - \left|1\right\rangle_2 \right) \right] \left|0\right\rangle_3 \\ + \left[ \left( \left|0\right\rangle_1 + \left|1\right\rangle_1 \right) \left( \left|0\right\rangle_2 - \left|1\right\rangle_2 \right) + \left( \left|0\right\rangle_1 - \left|1\right\rangle_1 \right) \left( \left|0\right\rangle_2 + \left|1\right\rangle_2 \right) \right] \left|1\right\rangle_3 \end{array} \right)$$

$$= \frac{1}{2} \left( \left( \left|00\right\rangle_{12} + \left|11\right\rangle_{12} \right) \left|0\right\rangle_3 + \left( \left|00\right\rangle_{12} - \left|11\right\rangle_{12} \right) \left|1\right\rangle_3 \right)$$

(11)

$$H_1 = H \otimes I \otimes I \qquad H_2 = I \otimes H \otimes I$$

By Equation (11), $q_1$ and $q_2$ are perfectly correlated when measured in the $\left\{ \left|+\right\rangle, \left|-\right\rangle \right\}$ basis. It is easy to verify that the same results apply to $\left| \psi^{(n)} \right\rangle$ such that any two qubits are not correlated in the $\left\{ \left|0\right\rangle, \left|1\right\rangle \right\}$ basis, but perfectly correlated in the $\left\{ \left|+\right\rangle, \left|-\right\rangle \right\}$ basis. Now if we apply the CHSH-test to the $q_1$-$q_2$ subsystem in $\left| \psi^{(3)} \right\rangle$, we have:

$$\text{For} \quad \left| \psi^{(3)} \right\rangle = \frac{1}{2} \left[ \left( \left|00\right\rangle_{12} + \left|11\right\rangle_{12} \right) \left|0\right\rangle_3 + \left( \left|01\right\rangle_{12} + \left|10\right\rangle_{12} \right) \left|1\right\rangle_3 \right],$$

$$S_{\max} \left( \left| \psi^{(3)} \right\rangle \right) = \left\langle A_0 \otimes B_0 \otimes I \right\rangle + \left\langle A_0 \otimes B_1 \otimes I \right\rangle - \left\langle A_1 \otimes B_0 \otimes I \right\rangle + \left\langle A_1 \otimes B_1 \otimes I \right\rangle = \sqrt{2}$$

(12)

In Equation (12), the maximum sum $S_{\max}$ has a reduced value of $\sqrt{2}$ as compared to the Bell states, indicating the dormant entanglement between $q_1$ and $q_2$ is weaker than the Bell states with reduced correlation. Note that the operators defined in Equation (8) measure $q_1$ and $q_2$ in a particular set of bases, and the $S_{\max}$ value in Equation (12) may change when the measurements are done in rotated bases. However, as details shown in the SI, the $S_{\max}$ value for $\left| \psi^{(3)} \right\rangle$ can never achieve $2\sqrt{2}$ like the Bell states, regardless of arbitrary basis rotations. This means that although the CHSH-test is dependent on basis rotations, it is still a useful indicator of the quantum correlation behaviors.

It turns out there can be a form of entanglement even weaker than the dormant entanglement of $q_1$ and $q_2$ in $\left| \psi^{(3)} \right\rangle$, consider the state:

$$\left| \psi^{(3+L)} \right\rangle = CX_{2 \to L} \left( \left| \psi^{(3)} \right\rangle \otimes \left|0\right\rangle_L \right)$$

$$= \frac{1}{2} \left[ \left( \left|00\right\rangle_{12} \left|0\right\rangle_L + \left|11\right\rangle_{12} \left|1\right\rangle_L \right) \left|0\right\rangle_3 + \left( \left|01\right\rangle_{12} \left|1\right\rangle_L + \left|10\right\rangle_{12} \left|0\right\rangle_L \right) \left|1\right\rangle_3 \right],$$

(13)

In Equation (13), $\left| \psi^{(3+L)} \right\rangle$ can be created by applying $CX_{2 \to L}$ to $\left( \left| \psi^{(3)} \right\rangle \otimes \left|0\right\rangle \right)_L$ to entangle the "lock" qubit $q_L$ to $q_2$. We have seen that $q_1$ and $q_2$ in $\left| \psi^{(3)} \right\rangle$ are not correlated in the $\left\{ \left|0\right\rangle, \left|1\right\rangle \right\}$

basis, and here the additional $q_L$ serves the role of locking this no-correlation property from getting changed by arbitrary basis transformations on $q_1$ and $q_2$:

$$U_1 U_2 \left| \psi^{(3+L)} \right\rangle = \frac{1}{2} \begin{pmatrix} \left[ \left( a_1 \left| 0 \right\rangle_1 + a_2 \left| 1 \right\rangle_1 \right) \left( b_1 \left| 0 \right\rangle_2 + b_2 \left| 1 \right\rangle_2 \right) \left| 0 \right\rangle_L + \left( a_2^* \left| 0 \right\rangle_1 - a_1^* \left| 1 \right\rangle_1 \right) \left( b_2^* \left| 0 \right\rangle_2 - b_1^* \left| 1 \right\rangle_2 \right) e^{i(\alpha+\beta)} \left| 1 \right\rangle_L \right] \left| 0 \right\rangle_3 \\ + \left[ \left( a_1 \left| 0 \right\rangle_1 + a_2 \left| 1 \right\rangle_1 \right) \left( b_2^* \left| 0 \right\rangle_2 - b_1^* \left| 1 \right\rangle_2 \right) e^{i\beta} \left| 1 \right\rangle_L + \left( a_2^* \left| 0 \right\rangle_1 - a_1^* \left| 1 \right\rangle_1 \right) \left( b_1 \left| 0 \right\rangle_2 + b_2 \left| 1 \right\rangle_2 \right) e^{i\alpha} \left| 0 \right\rangle_L \right] \left| 1 \right\rangle_3 \end{pmatrix}$$

$$U_1 = \begin{pmatrix} a_1 & a_2^* e^{i\alpha} \\ a_2 & -a_1^* e^{i\alpha} \end{pmatrix} \otimes I \otimes I \otimes I, \quad U_2 = I \otimes \begin{pmatrix} b_1 & b_2^* e^{i\beta} \\ b_2 & -b_1^* e^{i\beta} \end{pmatrix} \otimes I \otimes I, \quad \left| a_1 \right|^2 + \left| a_2 \right|^2 = \left| b_1 \right|^2 + \left| b_2 \right|^2 = 1$$

(14)

In Equation (14) $U_1$ and $U_2$ are arbitrary basis transformations applied to $q_1$ and $q_2$ respectively. Now without measuring $q_1$, the probability of measuring $q_2 = \left| 0 \right\rangle$ in $U_1 U_2 \left| \psi^{(3+L)} \right\rangle$ is:

$$p\left( q_2 = \left| 0 \right\rangle \right) = \frac{1}{4} \left[ \left( \left| a_1 \right|^2 + \left| a_2 \right|^2 \right) \left| b_1 \right|^2 + \left( \left| a_2^* \right|^2 + \left| a_1^* \right|^2 \right) \left| b_2^* \right|^2 + \left( \left| a_1 \right|^2 + \left| a_2 \right|^2 \right) \left| b_2^* \right|^2 + \left( \left| a_2^* \right|^2 + \left| a_1^* \right|^2 \right) \left| b_1 \right|^2 \right]$$

$$= \frac{1}{2}$$

(15)

If we measure $q_1$, suppose we get $q_1 = \left| 0 \right\rangle$, then $U_1 U_2 \left| \psi^{(3+L)} \right\rangle$ will collapse into the state:

$$U_1 U_2 \left| \psi^{(3+L)} \right\rangle \left( q_1 = \left| 0 \right\rangle \right) = \frac{1}{\sqrt{2}} \begin{pmatrix} \left[ a_1 \left| 0 \right\rangle_1 \left( b_1 \left| 0 \right\rangle_2 + b_2 \left| 1 \right\rangle_2 \right) \left| 0 \right\rangle_L + a_2^* \left| 0 \right\rangle_1 \left( b_2^* \left| 0 \right\rangle_2 - b_1^* \left| 1 \right\rangle_2 \right) e^{i(\alpha+\beta)} \left| 1 \right\rangle_L \right] \left| 0 \right\rangle_3 \\ + \left[ a_1 \left| 0 \right\rangle_1 \left( b_2^* \left| 0 \right\rangle_2 - b_1^* \left| 1 \right\rangle_2 \right) e^{i\beta} \left| 1 \right\rangle_L + a_2^* \left| 0 \right\rangle_1 \left( b_1 \left| 0 \right\rangle_2 + b_2 \left| 1 \right\rangle_2 \right) e^{i\alpha} \left| 0 \right\rangle_L \right] \left| 1 \right\rangle_3 \end{pmatrix}$$

(16)

And then the conditional probability of measuring $q_2 = \left| 0 \right\rangle$ given $q_1 = \left| 0 \right\rangle$ is:

$$p\left( q_2 = \left| 0 \right\rangle \mid q_1 = \left| 0 \right\rangle \right) = \frac{1}{2} \left( \left| a_1 b_1 \right|^2 + \left| a_2^* b_2^* \right|^2 + \left| a_1 b_2^* \right|^2 + \left| a_2^* b_1 \right|^2 \right) = \frac{1}{2} = p\left( q_2 = \left| 0 \right\rangle \right)$$

(17)

And therefore $q_1$ and $q_2$ are not correlated when measured in any arbitrary bases.

Equations (16) and (17) also help us to see why the lock qubit $q_L$ is important, because if we remove $q_L$ from Equation (16), we have:

$$U_1 U_2 \left| \psi^{(3)} \right\rangle \left( q_1 = \left| 0 \right\rangle \right) = \frac{1}{\sqrt{2}} \begin{pmatrix} \left[ a_1 \left| 0 \right\rangle_1 \left( b_1 \left| 0 \right\rangle_2 + b_2 \left| 1 \right\rangle_2 \right) + a_2^* \left| 0 \right\rangle_1 \left( b_2^* \left| 0 \right\rangle_2 - b_1^* \left| 1 \right\rangle_2 \right) e^{i(\alpha+\beta)} \right] \left| 0 \right\rangle_3 \\ + \left[ a_1 \left| 0 \right\rangle_1 \left( b_2^* \left| 0 \right\rangle_2 - b_1^* \left| 1 \right\rangle_2 \right) e^{i\beta} + a_2^* \left| 0 \right\rangle_1 \left( b_1 \left| 0 \right\rangle_2 + b_2 \left| 1 \right\rangle_2 \right) e^{i\alpha} \right] \left| 1 \right\rangle_3 \end{pmatrix}$$

$$(18)$$

And then the conditional probability of measuring $q_2 = \left| 0 \right\rangle$ given $q_1 = \left| 0 \right\rangle$ is:

$$
\begin{aligned}
p\left( q_2 = \left| 0 \right\rangle \mid q_1 = \left| 0 \right\rangle \right) &= \frac{1}{2} \left( \left| a_1 b_1 + a_2^* b_2^* e^{i(\alpha+\beta)} \right|^2 + \left| a_1 b_2^* e^{i(\beta)} + a_2^* b_1 e^{i(\alpha)} \right|^2 \right) \\
&= \frac{1}{2} \left( \left| a_1 b_1 \right|^2 + \left| a_2^* b_2^* \right|^2 + \left| a_1 b_2^* \right|^2 + \left| a_2^* b_1 \right|^2 \right) + \mathrm{Re}\left( a_1 b_1 a_2 b_2 e^{-i(\alpha+\beta)} \right) + \mathrm{Re}\left( a_1 b_1^* a_2 b_2^* e^{i(\beta-\alpha)} \right) \\
&= \frac{1}{2} + \mathrm{Re}\left( a_1 b_1 a_2 b_2 e^{-i(\alpha+\beta)} \right) + \mathrm{Re}\left( a_1 b_1^* a_2 b_2^* e^{i(\beta-\alpha)} \right) \\
&\neq p\left( q_2 = \left| 0 \right\rangle \right)
\end{aligned}
$$

$$(19)$$

We see that compared with Equation (16), in Equation (18) the lack of the lock qubit $q_L$ causes the term $a_1 \left| 0 \right\rangle_1 b_1 \left| 0 \right\rangle_2$ to mix with $a_2^* \left| 0 \right\rangle_1 b_2^* \left| 0 \right\rangle_2 e^{i(\alpha+\beta)}$, and $a_1 \left| 0 \right\rangle_1 b_2^* \left| 0 \right\rangle_2 e^{i(\beta)}$ to mix with $a_2^* \left| 0 \right\rangle_1 b_1 \left| 0 \right\rangle_2 e^{i(\alpha)}$, such that the conditional probability of $q_2 = \left| 0 \right\rangle$ given $q_1 = \left| 0 \right\rangle$ is the "square of the sum" in Equation (19) instead of the "sum of the square" in Equation (17). So without the lock qubit $q_L$, $q_1$ and $q_2$ can become correlated when measured in some rotated bases.

So indeed the entanglement of $q_1$ and $q_2$ in $\left| \psi^{(3+L)} \right\rangle$ is even weaker than that in $\left| \psi^{(3)} \right\rangle$ as there is no correlation in any basis. Now if we apply the CHSH-test to the $q_1$-$q_2$ subsystem in $\left| \psi^{(3+L)} \right\rangle$, we have:

For $\left| \psi^{(3+L)} \right\rangle = \frac{1}{2} \left[ \left( \left| 00 \right\rangle_{12} \left| 0 \right\rangle_L + \left| 11 \right\rangle_{12} \left| 1 \right\rangle_L \right) \left| 0 \right\rangle_3 + \left( \left| 01 \right\rangle_{12} \left| 1 \right\rangle_L + \left| 10 \right\rangle_{12} \left| 0 \right\rangle_L \right) \left| 1 \right\rangle_3 \right]$,

$$
\begin{aligned}
S_{\max}\left( \left| \psi^{(3+L)} \right\rangle \right) &= \pm \left\langle A_0 \otimes B_0 \otimes I \otimes I \right\rangle \pm \left\langle A_0 \otimes B_1 \otimes I \otimes I \right\rangle \pm \left\langle A_1 \otimes B_0 \otimes I \otimes I \right\rangle \pm \left\langle A_1 \otimes B_1 \otimes I \otimes I \right\rangle \\
&= 0
\end{aligned}
$$

$$(20)$$

In Equation (20), the sum $S$ is identically 0 with any combination of the signs before the observables, so $S_{\max} = 0$. In fact $S_{\max} = 0$ is the lowest value achievable for any quantum state, which can only happen if all four individual observables – $\left\langle A_0 \otimes B_0 \right\rangle$, $\left\langle A_0 \otimes B_1 \right\rangle$, $\left\langle A_1 \otimes B_0 \right\rangle$, and $\left\langle A_1 \otimes B_1 \right\rangle$ – in the CHSH-test are evaluated to zero (surprisingly, this zero value does not change when the observables are measured in arbitrary bases, see the SI for details).

Now is the entanglement of $q_1$ and $q_2$ in $\left|\psi^{(3+L)}\right\rangle$ a dormant entanglement as defined previously? Apply a Hadamard gate to $q_L$ in $\left|\psi^{(3+L)}\right\rangle$, we find:

$$
\begin{aligned}
H_L\left|\psi^{(3+L)}\right\rangle &= \frac{1}{2\sqrt{2}}\left[\begin{array}{l}\left(\left|00\right\rangle_{12}\left(\left|0\right\rangle_L+\left|1\right\rangle_L\right)+\left|11\right\rangle_{12}\left(\left|0\right\rangle_L-\left|1\right\rangle_L\right)\right)\left|0\right\rangle_3 \\ +\left(\left|01\right\rangle_{12}\left(\left|0\right\rangle_L-\left|1\right\rangle_L\right)+\left|10\right\rangle_{12}\left(\left|0\right\rangle_L+\left|1\right\rangle_L\right)\right)\left|1\right\rangle_3\end{array}\right] \\
&= \frac{1}{2\sqrt{2}}\left[\begin{array}{l}\left(\left(\left|00\right\rangle_{12}+\left|11\right\rangle_{12}\right)\left|0\right\rangle_L+\left(\left|00\right\rangle_{12}-\left|11\right\rangle_{12}\right)\left|1\right\rangle_L\right)\left|0\right\rangle_3 \\ +\left(\left(\left|01\right\rangle_{12}+\left|10\right\rangle_{12}\right)\left|0\right\rangle_L-\left(\left|01\right\rangle_{12}-\left|10\right\rangle_{12}\right)\left|1\right\rangle_L\right)\left|1\right\rangle_3\end{array}\right]
\end{aligned}
\tag{21}
$$

$H_L$ is the Hadamard gate applied on $q_L$

By Equation (21), the entanglement of $q_1$ and $q_2$ can be activated by measuring $q_3$ in the $\left\{\left|0\right\rangle,\left|1\right\rangle\right\}$ basis and $q_L$ in the $\left\{\left|+\right\rangle,\left|-\right\rangle\right\}$ basis, and then communicating the measurement results: $\left|+\right\rangle_L\left|0\right\rangle_3$ activates $\frac{1}{\sqrt{2}}\left(\left|00\right\rangle_{12}+\left|11\right\rangle_{12}\right)$, $\left|-\right\rangle_L\left|0\right\rangle_3$ activates $\frac{1}{\sqrt{2}}\left(\left|00\right\rangle_{12}-\left|11\right\rangle_{12}\right)$, $\left|+\right\rangle_L\left|1\right\rangle_3$ activates $\frac{1}{\sqrt{2}}\left(\left|01\right\rangle_{12}+\left|10\right\rangle_{12}\right)$, $\left|-\right\rangle_L\left|0\right\rangle_3$ activates $\frac{1}{\sqrt{2}}\left(\left|01\right\rangle_{12}-\left|10\right\rangle_{12}\right)$. If $q_3$ is measured in the $\left\{\left|+\right\rangle,\left|-\right\rangle\right\}$ basis, or $q_L$ is measured in the $\left\{\left|0\right\rangle,\left|1\right\rangle\right\}$ basis, then the entanglement of $q_1$ and $q_2$ is destroyed. Therefore indeed the entanglement of $q_1$ and $q_2$ is a form of dormant entanglement with multiple controllers, and all the controllers must reach complete consensus for measuring in the correct basis in order to activate the dormant entanglement.

Now by considering the correlation behavior in different bases and the CHSH-test results, we have three levels of entanglement:

Type 1: The Bell states: having perfect correlations in one basis and the Hadamard-rotated bases, and $S_{\max}=2\sqrt{2}$ for the CHSH-test (largest possible $S_{\max}$ due to the Tsirelson bound [34, 45]).

Type 2: The dormant entanglement of $\left|\psi^{(n)}\right\rangle$: having no correlation in one basis, but perfect correlation in the Hadamard-rotated basis, and $0<S_{\max}<2\sqrt{2}$ for the CHSH-test.

Type 3: The dormant entanglement of $\left|\psi^{(3+L)}\right\rangle$: having no correlation in any basis, and $S_{\max}=0$ for the CHSH-test (smallest possible).

We see that the dormant entanglement of $\left|\psi^{(3+L)}\right\rangle$ is the weakest possible form of entanglement, while the Bell states are the strongest.

## 5. Conclusion

In this work we first introduced the concept of "dormant entanglement" such that the entanglement of a subsystem can be activated or destroyed by measuring an external qubit in different bases. The fact that the subsystem can by no means determine the existence of the external system led to the unique quantum phenomenon that the physical description of a local system remains incomplete until information on all external systems entangled with the local system becomes available. We then extended the definition of dormant entanglement from the 3-qubit example to the general $n$-qubit system, and find that the structure of the dormant entanglement stays the same upon any permutation of the qubits – this then allowed us to propose a collective quantum communication channel in which any two parties can communicate through an activated entanglement channel with the complete consensus of all other parties. Finally we studied the correlation of the qubits when measured in different bases, with the help of the CHSH-test to quantify the correlation behavior. We found that by adding a "lock" qubit to the original dormant entanglement structure, a weakest form of entanglement was created such that it has no correlation in any basis and zero value (smallest possible) for the CHSH-test. For the future direction, all the results in this study were obtained by exact mathematical reasoning and therefore can be realized on any quantum computing hardware – photonic quantum systems are preferred due to their long-distance capability and resistance to noises and errors.

## 6. Acknowledgements

ZH and SK acknowledge funding by the U.S. Department of Energy (Office of Basic Energy Sciences) under Award No. DE-SC0019215; and the National Science Foundation under Award 2124511, CCI Phase I: NSF Center for Quantum Dynamics on Modular Quantum Devices (CQDMQD).

## 7. Supplementary Information

Supplementary Information is available after the **References**.

**References:**

1. Georgescu, I.M., S. Ashhab, and F. Nori, *Quantum simulation.* Reviews of Modern Physics, 2014. **86**(1): p. 153-185.
2. Montanaro, A., *Quantum algorithms: an overview.* npj Quantum Information, 2016. **2**(1): p. 15023.
3. Cao, Y., et al., *Quantum Chemistry in the Age of Quantum Computing.* Chemical Reviews, 2019. **119**(19): p. 10856-10915.
4. Albash, T. and D.A. Lidar, *Adiabatic quantum computation.* Reviews of Modern Physics, 2018. **90**(1): p. 015002.
5. Preskill, J., *Quantum Computing in the NISQ era and beyond.* Quantum, 2018. **2**: p. 79.

6.      Kais, S., ed. *Quantum Information and Computation for Chemistry*. Quantum Information and Computation for Chemistry. 2014, John Wiley & Sons.

7.      Preskill, J., *Quantum computing 40 years later.* arXiv:2106.10522 [quant-ph], 2021.

8.      Arute, F., et al., *Quantum supremacy using a programmable superconducting processor.* Nature, 2019. **574**(7779): p. 505-510.

9.      Boixo, S., et al., *Evidence for quantum annealing with more than one hundred qubits.* Nature Physics, 2014. **10**(3): p. 218-224.

10.     Linke, N.M., et al., *Experimental comparison of two quantum computing architectures.* Proceedings of the National Academy of Sciences, 2017. **114**(13): p. 3305.

11.     Carolan, J., et al., *Universal linear optics.* Science, 2015. **349**(6249): p. 711.

12.     Zhong, H.-S., et al., *Quantum computational advantage using photons.* Science, 2020. **370**(6523): p. 1460.

13.     Gong, M., et al., *Quantum walks on a programmable two-dimensional 62-qubit superconducting processor.* Science, 2021. **372**(6545): p. 948.

14.     Shor, P.W., *Polynomial-Time Algorithms for Prime Factorization and Discrete Logarithms on a Quantum Computer.* SIAM J. Comput., 1997. **26**(5): p. 1484–1509.

15.     Harrow, A.W., A. Hassidim, and S. Lloyd, *Quantum Algorithm for Linear Systems of Equations.* Physical Review Letters, 2009. **103**(15): p. 150502.

16.     Peruzzo, A., et al., *A variational eigenvalue solver on a photonic quantum processor.* Nature Communications, 2014. **5**(1): p. 4213.

17.     Daskin, A. and S. Kais, *Decomposition of unitary matrices for finding quantum circuits: Application to molecular Hamiltonians.* The Journal of Chemical Physics, 2011. **134**(14): p. 144112.

18.     Biamonte, J., et al., *Quantum machine learning.* Nature, 2017. **549**(7671): p. 195-202.

19.     Xia, R. and S. Kais, *Quantum machine learning for electronic structure calculations.* Nature Communications, 2018. **9**(1): p. 4195.

20.     Sajjan, M., S.H. Sureshbabu, and S. Kais, *Quantum Machine-Learning for Eigenstate Filtration in Two-Dimensional Materials.* Journal of the American Chemical Society, 2021. **143**(44): p. 18426-18445.

21.     Hu, Z., R. Xia, and S. Kais, *A quantum algorithm for evolving open quantum dynamics on quantum computing devices.* Scientific Reports, 2020. **10**(1): p. 3301.

22.     Wang, H., S. Ashhab, and F. Nori, *Quantum algorithm for simulating the dynamics of an open quantum system.* Physical Review A, 2011. **83**(6): p. 062317.

23.     Hu, Z., et al., *A general quantum algorithm for open quantum dynamics demonstrated with the Fenna-Matthews-Olson complex dynamics.* arXiv:2101.05287, 2021.

24.     Schlimgen, A.W., et al., *Quantum Simulation of Open Quantum Systems Using a Unitary Decomposition of Operators.* Physical Review Letters, 2021. **127**(27): p. 270503.

25.     Hu, Z. and S. Kais, *The unitary dependence theory for characterizing quantum circuits and states.* Communications Physics, 2023. **6**: p. 68.

26.     Hu, Z. and S. Kais, *The Quantum Condition Space.* Advanced Quantum Technologies, 2022. **5**(4): p. 2100158.

27.     Hu, Z. and S. Kais, *The wave-particle duality of the qudit quantum space and the quantum wave gates.* arXiv:2207.05213, 2022.

28.     Smart, S.E., et al., *Relaxation of stationary states on a quantum computer yields a unique spectroscopic fingerprint of the computer's noise.* Communications Physics, 2022. **5**(1): p. 28.

29. Hu, Z. and S. Kais, *Characterization of Quantum States Based on Creation Complexity.* Advanced Quantum Technologies, 2020. **3**(9): p. 2000043.

30. Hu, Z. and S. Kais, *A quantum encryption design featuring confusion, diffusion, and mode of operation.* Scientific Reports, 2021. **11**(1): p. 23774.

31. Hu, Z. and S. Kais, *Characterizing quantum circuits with qubit functional configurations.* Scientific Reports, 2023. **13**(1): p. 5539.

32. Bell, J.S., *On the Einstein Podolsky Rosen paradox.* Physics Physique Fizika, 1964. **1**(3): p. 195-200.

33. Pan, J.-W., et al., *Experimental test of quantum nonlocality in three-photon Greenberger–Horne–Zeilinger entanglement.* Nature, 2000. **403**(6769): p. 515-519.

34. Nielsen, M.A. and I.L. Chuang, *Quantum Computation and Quantum Information: 10th Anniversary Edition.* 2011: Cambridge University Press. 708.

35. Wang, X.-L., et al., *Quantum teleportation of multiple degrees of freedom of a single photon.* Nature, 2015. **518**(7540): p. 516-519.

36. Bennett, C.H. and G. Brassard, *Quantum cryptography: Public key distribution and coin tossing.* Theoretical Computer Science, 2014. **560**: p. 7-11.

37. Ekert, A.K., *Quantum cryptography based on Bell's theorem.* Physical Review Letters, 1991. **67**(6): p. 661-663.

38. Bennett, C.H., G. Brassard, and N.D. Mermin, *Quantum cryptography without Bell's theorem.* Physical Review Letters, 1992. **68**(5): p. 557-559.

39. Jennewein, T., et al., *Quantum Cryptography with Entangled Photons.* Physical Review Letters, 2000. **84**(20): p. 4729-4732.

40. Xu, F., et al., *Secure quantum key distribution with realistic devices.* Reviews of Modern Physics, 2020. **92**(2): p. 025002.

41. Yin, J., et al., *Entanglement-based secure quantum cryptography over 1,120 kilometres.* Nature, 2020. **582**(7813): p. 501-505.

42. Clauser, J.F., et al., *Proposed Experiment to Test Local Hidden-Variable Theories.* Physical Review Letters, 1969. **23**(15): p. 880-884.

43. Peres, A., *Separability Criterion for Density Matrices.* Physical Review Letters, 1996. **77**(8): p. 1413-1415.

44. Horodecki, M., P. Horodecki, and R. Horodecki, *Separability of mixed states: necessary and sufficient conditions.* Physics Letters A, 1996. **223**(1): p. 1-8.

45. Cirel'son, B.S., *Quantum generalizations of Bell's inequality.* Letters in Mathematical Physics, 1980. **4**(2): p. 93-100.

# Supplementary Information: Dormant entanglement that can be activated or destroyed by the basis choice of measurements on an external system

Zixuan Hu and Sabre Kais*

*Department of Chemistry, Department of Physics, and Purdue Quantum Science and Engineering Institute, Purdue University, West Lafayette, IN 47907, United States*
*\*Email:* kais@purdue.edu

## S1. The structure of $\left|\psi^{(n)}\right\rangle$ remains unchanged upon any permutation of the qubits.

In the main text Equation (6), we have shown the structure of $\left|\psi^{(3)}\right\rangle$ remains unchanged upon any permutation of the qubits. Now suppose the same applies to $\left|\psi^{(k)}\right\rangle$, if we can show it then also applies to $\left|\psi^{(k+1)}\right\rangle$, we can prove by mathematical induction that the structure of $\left|\psi^{(n)}\right\rangle$ remains unchanged upon any permutation of the qubits. To show this, we examine the creation procedure in Figure 2 of the main text, and recognize that:

$$\begin{aligned} \left|\psi^{(k+1)}\right\rangle &= \frac{1}{\sqrt{2}}\left(\left|\psi^{(k)}\right\rangle|0\rangle_{k+1} + \left|\tilde{\psi}^{(k)}\right\rangle|1\rangle_{k+1}\right) \\ &= \frac{1}{2}\left[\left(\left|\psi^{(k-1)}\right\rangle|0\rangle_k + \left|\tilde{\psi}^{(k-1)}\right\rangle|1\rangle_k\right)|0\rangle_{k+1} + \left(\left|\psi^{(k-1)}\right\rangle|1\rangle_k + \left|\tilde{\psi}^{(k-1)}\right\rangle|0\rangle_k\right)|1\rangle_{k+1}\right] \end{aligned} \qquad \text{S(1)}$$

In Equation S(1), $\left|\tilde{\psi}^{(k)}\right\rangle$ is defined as:

$$\left|\tilde{\psi}^{(k)}\right\rangle = X_k\left|\psi^{(k)}\right\rangle \qquad \text{S(2)}$$

such that $\left|\tilde{\psi}^{(k)}\right\rangle$ is the same as $\left|\psi^{(k)}\right\rangle$ except for the value of $q_k$ being flipped by the $X$ gate. Then by Equation S(1) we have:

$$\begin{aligned} \left|\psi^{(k+1)}\right\rangle &= \frac{1}{2}\left[\left(\left|\psi^{(k-1)}\right\rangle|0\rangle_k + \left|\tilde{\psi}^{(k-1)}\right\rangle|1\rangle_k\right)|0\rangle_{k+1} + \left(\left|\psi^{(k-1)}\right\rangle|1\rangle_k + \left|\tilde{\psi}^{(k-1)}\right\rangle|0\rangle_k\right)|1\rangle_{k+1}\right] \\ &= \frac{1}{2}\left[\left(\left|\psi^{(k-1)}\right\rangle|0\rangle_{k+1} + \left|\tilde{\psi}^{(k-1)}\right\rangle|1\rangle_{k+1}\right)|0\rangle_k + \left(\left|\psi^{(k-1)}\right\rangle|1\rangle_{k+1} + \left|\tilde{\psi}^{(k-1)}\right\rangle|0\rangle_{k+1}\right)|1\rangle_k\right] \end{aligned} \qquad \text{S(3)}$$

and thus the structure of $\left|\psi^{(k+1)}\right\rangle$ is unchanged when $q_k$ and $q_{k+1}$ are swapped. Now we have assumed that the structure of $\left|\psi^{(k)}\right\rangle$ is unchanged upon any qubit permutation, so indeed $q_k$ can be any qubit $q_j$ with $j = 1,...,k$ and the structure is unchanged upon swapping $q_j$ and $q_{k+1}$. In addition, once $q_j$ and $q_{k+1}$ have been swapped, $q_{k+1}$ becomes part of the $\left|\psi^{(k)}\right\rangle$ structure and thus can permute with any other qubits inside $\left|\psi^{(k)}\right\rangle$ without changing its structure. So we have the structure of $\left|\psi^{(k+1)}\right\rangle$ is unchanged upon any qubit permutation. Then by mathematical induction, we have proven the structure of a general $n$-qubit dormant entanglement state $\left|\psi^{(n)}\right\rangle$ remains unchanged upon any permutation of the qubits. In other words, all qubits in $\left|\psi^{(n)}\right\rangle$ are equivalent, therefore any two qubits are in dormant entanglement with the other qubits being its controllers.

## S2. Further discussions on the CHSH-test of $q_1$ and $q_2$ in $\left|\psi^{(3)}\right\rangle$ and $\left|\psi^{(3+L)}\right\rangle$.

In the main text Equation (12) we have $S_{\max}\left(\left|\psi^{(3)}\right\rangle\right)=\sqrt{2}$ for the observables defined in the main text Equation (8). Now if we rotate the CHSH observables, the $S_{\max}$ value may change, but it can never achieve $2\sqrt{2}$ like the Bell states – thus the CHSH-test still indicates that $\left|\psi^{(3)}\right\rangle$ has weaker correlations than the Bell states. To prove this, consider an observable $O$ as constructed in the CHSH style:

$$O = \pm A_0 \otimes B_0 \pm A_0 \otimes B_1 \pm A_1 \otimes B_0 \pm A_1 \otimes B_1$$

$$q_1 \text{ measurements: } \quad A_0 = U\sigma_z U^\dagger, \quad A_1 = U\sigma_x U^\dagger, \tag{S(4)}$$

$$q_2 \text{ measurements: } \quad B_0 = -\frac{1}{\sqrt{2}}V\left(\sigma_x + \sigma_z\right)V^\dagger, \quad B_1 = \frac{1}{\sqrt{2}}V\left(\sigma_x - \sigma_z\right)V^\dagger$$

here $A_0 = U\sigma_z U^\dagger$ and $A_1 = U\sigma_x U^\dagger$ for some arbitrary basis rotation $U$;

$B_0 = -\frac{1}{\sqrt{2}}V\left(\sigma_x + \sigma_z\right)V^\dagger$ and $B_1 = \frac{1}{\sqrt{2}}V\left(\sigma_x - \sigma_z\right)V^\dagger$ for some arbitrary basis rotation $V$; the signs before the $A_i \otimes B_j$ terms have to satisfy the CHSH condition: i.e. either the 3rd or the 4th sign has to be opposite to the other three signs.

Now we evaluate $O$ on the $q_1$-$q_2$ subsystem of $\left|\psi^{(3)}\right\rangle$ (the identity is applied to $q_3$):

$$S = \left\langle\psi^{(3)}\left|O \otimes I\right|\psi^{(3)}\right\rangle = \frac{1}{2}\left(\left\langle\varphi_1\left|\hat{O}\right|\varphi_1\right\rangle + \left\langle\varphi_2\left|\hat{O}\right|\varphi_2\right\rangle\right)$$

$$|\varphi_1\rangle = \frac{1}{\sqrt{2}}\left(|00\rangle_{12} + |11\rangle_{12}\right), \quad |\varphi_2\rangle = \frac{1}{\sqrt{2}}\left(|01\rangle_{12} + |10\rangle_{12}\right) \tag{S(5)}$$

So the expectation $S = \left\langle\psi^{(3)}\left|O \otimes I\right|\psi^{(3)}\right\rangle$ is the average of $\left\langle\varphi_1\left|\hat{O}\right|\varphi_1\right\rangle$ and $\left\langle\varphi_2\left|\hat{O}\right|\varphi_2\right\rangle$. Now by the Tsirelson bound [34, 45], the largest values for both $\left\langle\varphi_1\left|\hat{O}\right|\varphi_1\right\rangle$ and $\left\langle\varphi_2\left|\hat{O}\right|\varphi_2\right\rangle$ are the same $2\sqrt{2}$, but this max value cannot be achieved simultaneously for $\left\langle\varphi_1\left|\hat{O}\right|\varphi_1\right\rangle$ and $\left\langle\varphi_2\left|\hat{O}\right|\varphi_2\right\rangle$ – this means $\left\langle\psi^{(3)}\left|O \otimes I\right|\psi^{(3)}\right\rangle$ can never achieve $2\sqrt{2}$ regardless of the rotations $U$ and $V$. So although the $S_{\max}$ for $\left|\psi^{(3)}\right\rangle$ can change when the CHSH observables are rotated, it can never achieve $2\sqrt{2}$, and this means that $\left|\psi^{(3)}\right\rangle$ has reduced correlations as compared to the Bell states.

In the main text Equation (20), we have $S_{\max} = 0$ for $\left|\psi^{(3+L)}\right\rangle$ with the observables defined in the main text Equation (8). Here we prove that $S_{\max} = 0$ is always true regardless of arbitrary basis rotations on the observables. We use the $A$ and $B$ observables defined in Equation S(4) and then evaluate the four individual observables $\left\langle A_0 \otimes B_0\right\rangle$, $\left\langle A_0 \otimes B_1\right\rangle$, $\left\langle A_1 \otimes B_0\right\rangle$, and $\left\langle A_1 \otimes B_1\right\rangle$ separately. The algebra involved is quite long but straightforward, so here we only show brief steps for $\langle A_0 \otimes B_0 \rangle$:

$$\langle \psi^{(3+L)} | A_0 \otimes B_0 \otimes I \otimes I | \psi^{(3+L)} \rangle$$

$$= \frac{1}{4} \left( \langle \varphi_1 | A_0 \otimes B_0 | \varphi_1 \rangle + \langle \varphi_2 | A_0 \otimes B_0 | \varphi_2 \rangle + \langle \varphi_3 | A_0 \otimes B_0 | \varphi_3 \rangle + \langle \varphi_4 | A_0 \otimes B_0 | \varphi_4 \rangle \right)$$

$$|\varphi_1\rangle = \frac{1}{\sqrt{2}} \left( |00\rangle_{12} + |11\rangle_{12} \right), \ |\varphi_2\rangle = \frac{1}{\sqrt{2}} \left( |01\rangle_{12} + |10\rangle_{12} \right), \ |\varphi_3\rangle = \frac{1}{\sqrt{2}} \left( |01\rangle_{12} - |10\rangle_{12} \right), \ |\varphi_4\rangle = \frac{1}{\sqrt{2}} \left( |00\rangle_{12} - |11\rangle_{12} \right)$$

$$\text{S(6)}$$

This after some algebra becomes:

$$\langle \psi^{(3+L)} | A_0 \otimes B_0 \otimes I \otimes I | \psi^{(3+L)} \rangle$$

$$= \frac{1}{4} \left( \langle 00 | A_0 \otimes B_0 | 00 \rangle + \langle 11 | A_0 \otimes B_0 | 11 \rangle + \langle 01 | A_0 \otimes B_0 | 01 \rangle + \langle 10 | A_0 \otimes B_0 | 10 \rangle \right)$$

$$= \frac{1}{4} \left( \langle 0 | A_0 | 0 \rangle + \langle 1 | A_0 | 1 \rangle \right) \cdot \left( \langle 0 | B_0 | 0 \rangle + \langle 1 | B_0 | 1 \rangle \right)$$

$$= 0$$

$$\text{S(7)}$$

Note the result is always 0 regardless of the arbitrary basis rotations in S(4). This result also applies to $\langle A_0 \otimes B_1 \rangle$, $\langle A_1 \otimes B_0 \rangle$, and $\langle A_1 \otimes B_1 \rangle$: so all these four observables evaluate to zero individually, and thus $S_{max} = 0$ regardless of basis rotations for $|\psi^{(3+L)}\rangle$. $S_{max} = 0$ is indeed the lowest value achievable for any quantum state, indicating that the $q_1$-$q_2$ subsystem in $|\psi^{(3+L)}\rangle$ is the weakest form of entanglement with no correlation in any basis.